\documentclass[11pt]{article}

\newtheorem{theorem}{Theorem}
\newtheorem{lemma}{Lemma}
\newtheorem{corollary}{Corollary}
\newtheorem{remark}{Remark}
\newtheorem{proposition}{Proposition}

\makeatletter

\@addtoreset{equation}{section}
\makeatother

\def\ds{\displaystyle}
\def\e{\varepsilon}

\def\ti{\tilde}
\def\d{\partial}
\def\la{\langle}
\def\ra{\rangle}
\def\lla{\bigg\langle}
\def\rra{\bigg\rangle}

\def\H{{\cal H}}
\def\fH{\mbox{H}}
\def\fA{\mbox{A}}
\def\bxm{\mbox{\boldmath$\xi^{(\mu)}$}}

\def\bx1{\mbox{\boldmath$\xi^{(1)}$}}

\def\bJ{\mbox{\boldmath$J$}}
\def\bh{\mbox{\boldmath$h$}}
\def\xmo{\xi^{(\mu)}_1}
\def\xmi{\xi^{(\mu)}_i}

\def\xmj{\xi^{(\mu)}_j}

\def\no{\noindent}

\def\D{{\cal D}}
\def\P{\mbox{Prob}}

\def\a{\alpha}
\def\l{\lambda}
\def\dq{\dot q}

\def\bJm{{\bf J}^{-}}
\def\Smm{S^{-}_\mu}

\def\PP{{\cal P}}

\begin{document}
\begin{title}{Central limit theorems for order parameters
 of the Gardner problem}
\author{ M.Shcherbina\thanks{Institute for Low Temperature Physics,Ukr. Ac.
Sci., 47 Lenin ave., Kharkov, Ukraine}, B.Tirozzi\thanks{Department of Physics of
  Rome University "La Sapienza", 5, p-za A.Moro, Rome, Italy}}
\end{title}
\date{}
\maketitle

\begin{abstract}
Fluctuations of the order parameters of the Gardner
model for any $\alpha<\alpha_c$ are studied. It is proved that they
converge in distribution to a family of jointly Gaussian random
variables.
\end{abstract}

\section{Introduction and Main Results}
The Gardner model was introduced in \cite{G} to study the typical
volume of interactions between each pair of $N$ Ising spins which
solve the problem of storing a given set of $p$ random patterns
$\{\bxm\}_{\mu=1}^p$. The components $\xmi$ of the patterns are
 taken usually to be independent random variables with zero mean
 and variance 1.
After a simple transformation this problem is reduced to
the analysis  of the asymptotic behaviour of the random variable
\begin{equation}
\Theta_{N,p}(k)= \sigma_N^{-1}\int_{(\bJ,\bJ)= N} d\bJ\prod_{\mu=
1}^p \theta(N^{-1/2}(\bxm,\bJ)-k),
\label{Theta}\end{equation}
where the function $\theta(x)$, as usually, is zero in the
negative semi-axis and 1 in the positive and  $\sigma_N$ is the
Lebesgue measure of $N$-dimensional sphere of radius $N^{1/2}$.
Then, the question of interest is the behaviour of
$\frac{1}{N}\log\Theta_{N,p}(k)$ in the limit $N,p\to \infty$,
$\frac{p}{N}\to\alpha$. Gardner \cite{G} had solved this problem
by using the so-called replica trick, which is completely
non-rigorous from the  mathematical point of view but sometimes
very useful in the  physics of spin glasses (see \cite{MPV} and
references therein).

She obtained  that for any $\alpha<\alpha_c(k)$, where
\begin{equation}
\a_c(k)\equiv ({1\over
\sqrt{2\pi}}\int_{-k}^\infty(u+k)^2e^{-u^2/2}du)^{-1},
\label{a_c}\end{equation}
the following limit exists
\begin{equation}\begin{array}{lll}\ds{
\lim_{N,p\to\infty,p/N\to\a}\frac{1}{N}E\{\log \Theta_{N,p}(k)\}}&=& \ds{
 \min_{0\le q\le 1}\left[\a
E\left\{\log \fH\left({u\sqrt
q+k\over\sqrt{1-q}}\right)\right\}\right.}\\
&&\ds{\left.+{1\over 2}{q\over 1-q}+{1\over
2}\log(1-q)\right],}
\end{array}\label{cal_F}\end{equation}
where $u$ is a Gaussian random variable with zero mean and
variance $1$, $\fH(x)$ is defined as
\begin{equation}
\fH(x)\equiv{1\over\sqrt{2\pi}}\int_x^\infty e^{-t^2/2}dt
\label{H}\end{equation}
 and here and below we denote by the symbol
$E\{...\}$ the averaging with respect to all random parameters of
the problem and also with respect to $u$. And
$\frac{1}{N}\log\Theta_{N,p}(k)$ tends to minus
infinity for $\alpha\ge\alpha_c(k)$.

In the paper \cite{ST3} (see also \cite{ST4}) we have studied the Gardner problem in a
regular mathematical way and proved that for any
$\alpha<\alpha_c$ formula (\ref{cal_F}) is valid while for
$\alpha>\alpha_c$ any $\frac{1}{N}E\{\log \Theta_{N,p}(k)\}\to
-\infty$, as $N,p\to\infty,p/N\to\a$. We studied the case
$\xmi=\pm1$, but of course the same results are valid for any
distribution $\xmi$, if $E|\xmi|^4<\infty$.

To obtain this results  we introduced an intermediate  "modified"
Hamiltonian depending on the parameters $\e>0$,  $z>0$.
\begin{equation}
\H(\bJ,k,h,z,\e)\equiv-\sum_{\mu= 1}^p\log \fH\left(
\frac{k-(\bxm,\bJ)N^{-1/2}}{\sqrt\e}\right)+({\bf h},\bJ)
+ {z\over 2}(\bJ,\bJ),
\label{H_N,p}\end{equation} where   the function
$\fH(x)$ is defined in (\ref{H})
 and $\bh= (h_1,...,h_N)$ is an external
random field with independent Gaussian $h_i$ with zero mean and
variance $1$.

The partition function for this Hamiltonian is
\begin{equation}
Z_{N,p}(k,h,z,\e)= \sigma_N^{-1}\int
d\bJ\exp\{-\H_\e(\bJ,k,h,z,\e)\}. \label{Z}\end{equation} We
denote also by $\la\dots\ra$ the corresponding Gibbs averaging and
\begin{equation}
f_{N,p}(k,h,z,\e)\equiv{1\over N}\log Z_{N,p}(k,h,z,\e).
\label{f}\end{equation}
In \cite{ST3} we have proved the theorem:
\begin{theorem}\label{thm:1}
For any $\alpha<2$, $k>0$, there exists $\e^*(\alpha,k)$ such that
for any $\e\le\e^*(\alpha,k)$  and $z\le\e^{-1/3}$  there exists
\begin{equation}\begin{array}{rll}\ds{
\lim_{N,p\to\infty,\a_N\to\a}E\{f_{N,p}(k,h,z,\e)\}}&=&
F(\a,k,h,z,\e),\\
 F(\a,k,h,z,\e)&\equiv& \ds{\max_{ R>0}\min_{0\le
q\le R}\left[\a E\left\{\log \emph{H}\left({u\sqrt
q+k\over\sqrt{\e+R-q}}\right)\right\} \right.}\\
&& \ds{
\left.+{1\over 2}{q\over R-q}+{1\over 2}\log(R-q)-{z\over 2}R
+{h^2\over 2}( R-q)\right],}
\end{array}\label{t1.1}\end{equation}
where  $u$ is a Gaussian random variable with zero mean and variance 1.
\end{theorem}
Similar results for small $\alpha$ were obtained in \cite{Tal2}
for the so-called Gardner-Derrida model. In the paper \cite{Tal4}
the fluctuation of the order parameters for the Gardner-Derrida
model were studied, but  only for small enough $\alpha$.

An important ingredient of the analysis of the free energy of the
model (\ref{H_N,p}) in \cite{ST3} was the proof of the fact that the  variance of
its  order parameters (or the overlap
parameters) disappears in the thermodynamic limit.
In the present paper we study the behaviour of  fluctuations of
the overlap parameters, defined as
\begin{equation}
 R_{l,m}=\frac{1}{N}(\bJ^{(l)},\bJ^{(m)}), \quad (l,m=1,\dots n),
\label{R_lm}\end{equation} where  the upper indexes of the
variables $\bJ$ mean that we consider  $n$ replicas of the
Hamiltonian (\ref{H_N,p}) with the same random parameters
$\{\bxm\}_{\mu=1}^p$, $\{{\bf h}\}$, but different
$\bJ^{(1)},\dots, \bJ^{(n)}$.

We introduce  also the notations:
\begin{equation}\begin{array}{ll}
  \dq=N^{1/2}(\la R_{1,2}\ra-q),&\\
\ds{
T_{l,m}=\frac{1}{N^{1/2}}(\dot{\bJ}^{(l)},\dot{\bJ}^{(m)}),}&\ds{
 T_l=\frac{1}{ N^{1/2} } (\dot{\bJ}^{(l)},\la\bJ\ra) .}
\end{array}\label{RT}\end{equation}
Here and below  $\dot{\bf J}\equiv {\bf J}-\la {\bf
J}\ra$ and
 $(q,R)$ is the unique solution of the system of equations:
\begin{equation}\begin{array}{lll}
 q&=&\ds{(R- q)^2\bigg[ {\alpha\over
{R-q+\e}}E\left\{\fA^2\left({\sqrt{ q}u+k\over \sqrt
{R-q+\e}}\right)\right\}+h^2\bigg],}\\
z&=& \ds{{\alpha \over
(R-q+\e)^{3/2}}E\left\{(\sqrt{ q}u+k) \fA\left({\sqrt{
q}u+k\over \sqrt {R-q+\e}}\right)\right\}}\\
& &\ds{
 -{q\over( R-q)^2}+ {1\over R-q} +h^2,}
\end{array}\label{q,R}\end{equation}
with
$$
\fA(x)=-\frac{1}{\sqrt{2\pi}}\frac{d}{dx}\log \fH(x).
$$
To avoid additional technical difficulties below we assume that
$\{\xmi\}$ are independent Gaussian random variables with zero
mean and variance 1.

The main result of the paper is
\begin{theorem}\label{thm:2} Consider
 any $\alpha<2$, $k>0$, $\e\le\e^*(\alpha,k)$
and $z\le\e^{-1/3}$. Then for any integer $n$
the families of random variables $\{\sqrt N ( R_{l,m}-E\la R_{l,m}\ra)\}_{l<m\le n}$,
 converges in distribution, as
$N,p\to\infty,p/N\to\a$, to the Gaussian family of random variables
$\{v_{l,m}\}_{l<m\le n}$,
 with the covariance matrix:
\begin{equation}\begin{array}{lll}
E\{v_{l,m}v_{l,m}\}&=&A_0,\\
E\{v_{l,m}v_{l,m'}\}&=&B_0\,\, (m\not=m'),\\
E\{v_{l,m}v_{l',m'}\}&=&C_0\,\, (m,m',l, l' \,\,are\,\,
different).
\end{array}\label{t2.1}\end{equation}
In particular,
\begin{equation}\begin{array}{lll}
\lim_{N,p\to\infty,p/N\to\a}E\left\{\la T_{1,2}^{2n}
\ra\right\}&=&\frac{\Gamma(2n-1)}{\Gamma(n-1)}
A^n\\
\lim_{N,p\to\infty,p/N\to\a}E\left\{\la T_{1}^{2n}
\ra\right\}&=&\frac{\Gamma(2n-1)}{\Gamma(n-1)}
B^n\\
\lim_{N,p\to\infty,p/N\to\a}E\{\dq^{2n}\}&=&
\frac{\Gamma(2n-1)}{\Gamma(n-1)}
C^n,
\end{array}\label{t2.0}\end{equation}
where the  constants $A_0$, $B_0$, $C_0$, $A$, $B$, $C$ depend on $\alpha,k,z,\e$ and
all odd moments for these random variables tend to zero.
\end{theorem}
\begin{remark} In fact it follows from our proof (see  proofs of Lemmas
3,4,5 in Sec.2) that $\{T_{l,m}\}_{l<m\le n}$ and $\{T_l\}_{l\le n}$
in some sense do not depend on the random variables $\{\xmi\}$ and
$\{h_i\}$, i.e. if we consider $P$- some product of
$\{T_{l,m}\}_{l<m\le n}$ and $\{T_l\}_{l\le n}$, then
\begin{equation}
\lim_{N,p\to\infty,p/N\to\a}E\{(\la P\ra-E\la P\ra)^2\}=0.
\end{equation}
\end{remark}

As it was mentioned above, similar results were obtained in
\cite{Tal4} for the Gardner-Derrida model for small $\alpha$ and
for the Sherrington-Kirkpatrick model for the high temperature.
We would like to mention also the work \cite{GuT}, where the
fluctuations of the overlap parameters for the Sherrington-Kirkpatrick
model in the high temperature region were studied by the method
of characteristic functions.

One of the most important feature of our Hamiltonian (\ref{H_N,p}),
which allows us to prove Theorems \ref{thm:1} and \ref{thm:2}
for any $\alpha<\alpha_c(k)$ is that
it has the form
\begin{equation}
-{\cal H}=\sum_\mu g(S_\mu)-\frac{z}{2}(\bJ,\bJ)-({\bf h,\bJ}),\quad
S_\mu=\frac{1}{N^{1/2}}(\bJ,\bxm),
\label{conc}\end{equation}
where $g(x)$ is a concave function. It allows us to use
 the Brascamp-Lieb  inequalities ( see \cite{BL}),  according to which
  for any integer $n$  and any  ${\bf x}\in {\bf R}^N$
\begin{equation}
\lla \bigg(\frac{(\dot{\bJ},{\bf x})}{\sqrt N}\bigg)^{2n}\rra
\le\frac{\Gamma(2n-1)}{z^{n}\Gamma(n-1)}\bigg(\frac{|{\bf
x}|^2}{N} \bigg)^n.
\label{bl.1}\end{equation}
 Besides,
 for any  smooth function $f$
\begin{equation}
\la (f-\la f\ra)^2\ra\le\frac{1}{z}\la |\nabla f|^2\ra.
\label{bl.2}\end{equation}
Below we  use the representation (\ref{conc}) of $\H$ and the following
 properties of the functions $g(x)$:
 \begin{equation}
g(x)\le 0,\,\,-C\le g''\le 0,\,\, |g^{(s)}(x)|\le C,\,\,
(s=3,\dots,6).
\label{cond_g}\end{equation}
In fact the only place where we use the real form of $g(x)$ is that the
limiting system of equations (\ref{q,R}) has the unique
solution (see \cite{ST3}).

We use also notations:
\begin{equation}
S_\mu^{(l)}=\frac{1}{N^{1/2}}(\bJ^{(l)},\bxm),\quad
\ti R_{l,l'}=\frac{1}{N}\sum_\mu
g'(S_\mu^{(l)})g'(S_\mu^{(l')}),\quad \ti U_{l}=\frac{1}{
N}\sum_\mu (g''(S_\mu^{(l)})+g'^2(S_\mu^{(l)})).
\label{ti_R}\end{equation}

An important ingredient of our proof is the following proposition:

\begin{proposition}\label{pro:1} There exists $d_0>0$ such that
for any $\delta<d_0$
\begin{equation}\begin{array}{l}\ds{
\emph{\P}\{|\la R_{1,2}\ra-E\la R_{1,2}\ra|>\delta\}\le e^{-N\delta^2/2C},}\\
\ds{
\emph{\P}\{|\la R_{1,1}\ra-E\la R_{1,1}\ra|>\delta\}\le e^{-N\delta^2/2C},}\\
\ds{
\emph{\P}\{|\la \ti R_{1,2}\ra-E\la \ti R_{1,2}\ra|>\delta\}\le e^{-N\delta^2/2C},}\\
\ds{
\emph{\P}\{|\la \ti U_1\ra-E\la \ti U_1\ra|>\delta\}\le e^{-N\delta^2/2C}.}
\end{array}\label{p1.1}\end{equation}
\end{proposition}

\begin{corollary}\label{cor:1}
\begin{equation}
E\{|\la R_{1,2}\ra-E\la R_{1,2}\ra|^n\}
\le 2\frac{C^n\Gamma(n)}{N^{n/2}}.
\label{p1.2}\end{equation}
\end{corollary}

\section{Proof of Main Results}

\no {\it Proof of Proposition \ref{pro:1}}

\no For the proof of  Proposition \ref{pro:1} we need the
following remark:
\begin{remark}
It was proven in \cite{ST3} that there exist
  constants $M_0$, $m_0$ such that for any $M>M_0$
\begin{equation}
\emph{\P}\{\la(\bJ,\bJ)\ra\ge MN\}\le e^{-Nm_0(M-M_0)}.
\label{bound_J}\end{equation}
Besides, it is well known that if we define
\begin{equation}
\mathcal{A}_{*i,j}=\frac{1}{N}\sum_\mu\xmi\xmj,
\label{A^*}\end{equation}
then there exist $C_0$, $c_0$ such that for any $C>C_0$
\begin{equation}
\P\{||\mathcal{A}_*||\ge C\}\le e^{-Nc_0(C-C_0)}
\label{bound_A}\end{equation}
(see e.g. \cite{ST1}).
\end{remark}

\no We prove Proposition \ref{pro:1} using a method proposed in \cite{Tal4}
with small modifications which we need to study the case, when
the variables $ \{J_i\}$ are unbounded.
\begin{lemma}\label{lem:1}
Consider two Gaussian independent random vectors
$\{{\bf u}\}$ and $\{{\bf v}\}$. Let $f(\bf x)$ satisfies the conditions:
\begin{equation}
 P(A)=\emph{\P}\{|\nabla f({\bf u}))|^2\ge A\}
\le e^{-C(A-A_0)}\,\,
(\forall A>A_0)
\label{p1.3}\end{equation}
and for some $ s_0$
\begin{equation}
E\{e^{\pm  s_0 f}\}\le e^{s_0B}.
\label{p1.3a}\end{equation}
Then for $s\le\frac{1}{4}s_0$
\begin{equation}
E\{e^{\pm s(f({\bf u})-Ef({\bf u}))}\}\le e^{ 2A_0s^2}(1+
(A_0+C^{-1})e^{2sB-CA_0/2}).
\label{p1.4}\end{equation}

\end{lemma}
\no {\it Proof.}

\no The proof is very simple.
Consider
$$
G_{s,t}({\bf u},{\bf v})= \exp\{s(f({\bf u})-f(\sqrt{1-t}{\bf
u}+\sqrt{t}{\bf v}))\}, \quad \varphi_s(t)=E\{G_{s,t}({\bf u},{\bf
v})\}.$$ Then, integrating by parts, we get
$$
\begin{array}{lll}
\varphi_s'(t)&=&\ds{\frac{s}{2}
E\left\{\frac{\d }{\d x_i}f(\sqrt{1-t}{\bf u}+\sqrt{t}{\bf v})
(\frac{u_i}{\sqrt{1-t}}-
\frac{v_i}{\sqrt{t}})G_{s,t}({\bf u},{\bf v})\right\} }\\
&=&\ds{\frac{s^2}{2\sqrt{1-t}}E\left\{\frac{\d }{\d x_i}f(\sqrt{1-t}{\bf u}+
\sqrt{t}{\bf v})
\frac{\d }{\d x_i}f({\bf u})G_{s,t}({\bf u},{\bf v})\right\}}\\
&\le &\ds{\frac{A_0s^2}{\sqrt{1-t}}\varphi_s(t)+
\frac{s^2e^{2sB}}{2\sqrt{1-t}}E^{1/2}\left\{|\nabla f (u)|^4
\theta(|\nabla f (u)|^2-2A_0)\right\}}\\
&\le &\ds{ \frac{A_0s^2}{\sqrt{1-t}}\varphi_s(t) +
\frac{s^2e^{2sB}}{2\sqrt{1-t}}\left(\int_{A>2A_0} A^2dP(A)\right)^{1/2}.}
\end{array}$$
Thus we have got
$$\begin{array}{l}
E\{e^{s(f({\bf u})-f({\bf v}))}\}=
\varphi_s(1)\le
e^{2A_0s^2}(\varphi_s(0)+(A_0+C^{-1})e^{2sB-CA_0/2}).
\end{array}$$
Since by definition $\varphi_s(0)=1$,
averaging first  with respect to ${\bf u}$ and then with respect to ${\bf v}$
and using the Iensen inequality, we get (\ref{p1.4}).
Lemma \ref{lem:1} is proven.

\smallskip

To apply this result to $f=N\la R_{1,2}\ra$ we have to check the condition
(\ref{p1.3}). We write
\begin{equation}\begin{array}{lll}\ds{
\frac{1}{N}\sum \bigg|\frac{\d f}{\d \xmi}\bigg|^2}&=&
\ds{{4\over N^2}\sum_{i,j,k,\mu}
\la J_i\ra\lla \dot J_i g'(S_\mu)J_j\rra \lla  J_j g'(S_\mu)\dot J_k\rra
\la J_k\ra}\\
&\le &\ds{{8\over N^2}\sum_{i,j,k,\mu}
\la J_i\ra\lla \dot J_i g'(S_\mu)\dot J_j\rra
\lla \dot J_j g'(S_\mu)\dot J_k\rra\la  J_k\ra}\\
&&\ds{+{8\over N^2}\sum_{i,j,k,\mu} \la  J_i\ra
\lla\dot J_i( g'(S_\mu)-\la g'(S_\mu)\ra)\rra
\lla (g'(S_\mu)-\la g'(S_\mu)\ra)\dot J_k\rra\la  J_k\ra\la J_j\ra^2}\\
&=&8I+8II.
\end{array}\label{p1.5}\end{equation}

To estimate the r.h.s. we use the following proposition:
\begin{proposition}\label{pro:2}
Consider the matrices  $\mathcal{A}^{(f)}:{\bf R}^N\to{\bf R}^N$,
$\mathcal{B}^{(f)}:{\bf R}^p\to{\bf R}^N$ and
$\mathcal{C}:{\bf R}^p\to{\bf R}^p$ of the form
\begin{equation}\begin{array}{l}
\mathcal{A}_{i,j}^{(f)}=\lla \dot J_i\dot J_j f(\bJ)\rra,\quad
\mathcal{B}_{i,\mu}^{(f)}=\lla \dot J_i
(g'(S_\mu)-\la g'(S_\mu)\ra) f(\bJ)\rra,\\
\mathcal{C}_{\mu,\nu}=\lla (g'(S_\mu)-\la g'(S_\mu)\ra)
(g'(S_\nu)-\la g'(S_\nu)\ra) \rra.
\end{array}\label{A,B}\end{equation}
Then
\begin{equation}
||\mathcal{A}^{(f)}||\le \frac{\la 3 f^2\ra^{1/2}}{z},\quad
||\mathcal{B}^{(f)}||\le \frac{||\mathcal{A}_*||^{1/2}\la|g''|^2\ra^{1/2}
\la 3f^4\ra^{1/4}}{z},\quad
||\mathcal{C}||\le \frac{||\mathcal{A}_*||\la|g''|^2\ra}{z},
\label{p2.1}\end{equation}
where the matrix $\mathcal{A}_*$ is defined by (\ref{A^*}).
\end{proposition}
We prove this proposition in the next section.

Denoting $\mathcal{A}^{\mu}_{i,j}=\la \dot J_i\dot J_j g'(S_\mu)\ra$
and using (\ref{p2.1}),  we obtain
$$
I=\frac{1}{N^2}\sum (\mathcal{A}^{\mu}*\mathcal{A}^{\mu}\la\bJ\ra,\la\bJ\ra)\le
\frac{3}{z^2N^2}(\la\bJ\ra,\la\bJ\ra)\lla\sum g'^2(S_\mu)\rra.
$$
Similarly, taking  $f=1$ in the definition (\ref{p2.1}) of $\mathcal{B}$,
we have got
$$
II=\frac{1}{N^2}(\la\bJ\ra,\la\bJ\ra)
(\mathcal{B}*\mathcal{B}^*\la\bJ\ra,\la\bJ\ra)\le
\frac{2||\mathcal{A}_*||\la|g''|^2\ra}{z^2N^2}(\la\bJ\ra,\la\bJ\ra)^2.
$$

\begin{proposition}\label{pro:3}
$$
\frac{1}{N}\sum\la g'^2(S_\mu)\ra\le \frac{C}{N^2}\sum(\bxm,\bxm).
$$
\end{proposition}
Since $g''$ is bounded function, by using this proposition
(see the next section for the proof) one can easily
check the condition (\ref{p1.3}) for the terms $I$ and $II$.

Thus we have proved the first line of (\ref{p1.4}). Now by using the standard Chebyshev
inequality we get (\ref{p1.1}). The other inequalities in (\ref{p1.1})
can be proven similarly.

\smallskip

To prove Theorem \ref{thm:2} we need to make some preliminary work.

\no Denote

\begin{equation}
d= q(R-q)^{-1},\quad U=d+z-(R-q)^{-1};
\label{d,U}\end{equation}
\begin{lemma}\label{lem:2} For any $0<\epsilon<1$ there exists  a
constant $C_\epsilon$ such that uniformly in $N$
\begin{equation}\begin{array}{ll}\ds{
|E\{\la R_{1,2}\ra\}-q|\le \frac{C_\epsilon}{N^{1-\epsilon}}, }&\ds{
|E\{\la R_{1,1}\ra\}-R|\le
\frac{C_\epsilon}{N^{1-\epsilon}}},
\\
\ds{|E\{\la g'(S_\mu)\ra^2\}-d|\le \frac{C_\epsilon}{N^{1-\epsilon}},
}&\ds{|E\{\la g''(S_\mu)+g'^2(S_\mu)\ra\}-U|\le
\frac{C_\epsilon}{N^{1-\epsilon}}}.
\end{array}\label{l2.1}\end{equation}
\end{lemma}
For the proof of this lemma see Section 3.

Using this lemma and inequality (\ref{p1.2}), we get
\begin{equation}
E\{|\dq^n|\}\le 2C^n\Gamma(n).
\label{t2.2}\end{equation}
Besides, using inequalities (\ref{bl.1}) one can get easily:
$$
E\{\la T_{1,2}^n\ra\}\le C^n\Gamma(n),\quad
E\{\la T_{1}^n\ra\}\le C^n\Gamma(n).
$$
Then, since $R_{1,2}-q=N^{-1/2}(T_{1,2}+T_1+T_2+\dq)$, we obtain
\begin{equation}
E\{|\la(R_{1,2}-q)^n\ra\}\le \frac{C^n\Gamma(n)}{N^{n/2}}.
\label{main1}\end{equation}
Besides,
on the basis of (\ref{bl.2}) and Lemma \ref{lem:2}, be have got
\begin{equation}
E\{\la(\ti R_{l,l'}-d)^2\ra\}\le\frac{C}{N},\quad
E\{\la\dot {U}_l^2\ra\}\le\frac{C}{N},\quad E\{\la( R_{l,l}-R)^2\ra\}
\le\frac{C}{N}.
\label{main2}\end{equation}
Here and below we denote
$$
\dot {U}_l=\ti{ U}_l-U.
$$
From this inequality, using  the bound
$$
|\ti R_{l,l'}|,|\dot U^2|\le N^{-1}\la(\bJ,\bJ)\ra||\mathcal{A}_*||
$$
and  inequalities (\ref{bound_J}) and (\ref{bound_A}),
we obtain for any $r>2$
\begin{equation}
E\left\{\la|\ti R_{l,l'}-d|^r\ra\right\}\le\frac{C_r}{N},\quad
E\left\{\la|\dot U|^r\ra\right\}\le\frac{C_r}{N},\quad
E\left\{\la|R_{l,l}-R|^r\ra\right\}\le\frac{C_r}{N}.
\label{main3}\end{equation}

Following the method of \cite{Tal5}, we introduce the Hamiltonian
\begin{equation}\begin{array}{lll}
-H_t&=&\ds{\sum g(\Smm+J_1\sqrt t \xmo N^{-1/2} )
+\sqrt{d(1-t)}uJ_1}\\
&& \ds{
 +\frac{1-t}{2}(U-d)J_1^2-\frac{z}{2}J_1^2-\frac{z}{2}(\bJm,\bJm),}
\end{array}\label{H(t)}\end{equation}
where $u$ is a normally distributed random variable, independent
of $\bxm$ and ${\bf h}$
 and $\Smm=N^{-1/2}(\bxm,\bJm)$ do not depend on $\xmo$.

Denote $\la...\ra_t$ the Gibbs averaging to $H_t$
(or $n$ replicas of $H_t$), and for any $\xmo$-independent
function defined of ${\bf R}^{N\times n}$ define
\begin{equation}
\nu_t(f)=E\la f\ra_t, \quad\nu_t'(f)=\frac{d}{dt}\nu_t(f).
\label{nu_t}\end{equation}
for any $\xmo$-independent
function defined of ${\bf R}^{N\times n}$.
Besides, to simplify notation we  denote
$$
s_i=J^{(i)}_1.
$$

\begin{proposition}\label{pro:4}
 For any integer $n$
\begin{equation}
|\nu_t(s_1^{2n})|\le C^n\Gamma(n).
\label{p5.1}\end{equation}
\end{proposition}
For the proof of this proposition see the next section.
\smallskip

Let us compute $\nu_t'(f)$. Differentiating and then integrating by parts
with respect to $\xmo$ and $u$, we have got
\begin{equation}\begin{array}{lll}
\nu_t'(f)&=&\ds{\frac{1}{2}\sum_{l=1}^n \nu_t(f s_l^2\dot U_l^{-})-
\frac{n}{2}\nu_t(fs_{n+1}^2\dot U_{n+1}^{-})}\\
&&\ds{
+\sum_{l<l'}^n\nu_t(fs_ls_{l'}
(\ti R_{l,l'}^{-}-d))
-n\sum_{l=1}^n\nu_t(fs_ls_{n+1}(\ti R_{l,n+1}^{-}-d))}\\
&&\ds{
+\frac{n(n+1)}{2}\nu_t(fs_{n+1}s_{n+2}(\ti R_{n+1,n+2}^{-}-d))}.
\end{array}\label{nu'}\end{equation}
 Since the Hamiltonian (\ref{H(t)})
has the form (\ref{conc}), the inequalities (\ref{bl.1}) and (\ref{bl.2})
for this Hamiltonian  are also valid. Therefore the estimate
(\ref{main1}), and
(\ref{main3}) are fulfilled and so, using the Schwartz inequality and
(\ref{main3}), we get
\begin{equation}
|\nu_1(f)-\nu_0(f)|\le C\max_t\nu_t^{1/2}(|f|^2) N^{-1/2}.
\label{main4}\end{equation} Using the same formula to compute the
second derivative of $\nu_t(f)$ with respect to $t$, we obtain the
expression in each term of which we have $(\ti R_{l,l'}-d)(\ti
R_{l_1,l'_1}-d)$ or $(\ti R_{l,l'}-d)\dot U_{l_1}$ or $\dot
U_{l}\dot U_{l_1}$. Using again (\ref{main3}) and the H\"{o}lder
inequality, we obtain for any $0<\epsilon<1$:
\begin{equation}
|\nu_1(f)-\nu_0(f)-\nu_0'(f)|\le \max_t\nu_t^{\epsilon/2}(|f|^{2/\epsilon})
\nu_t^{\epsilon/2}(|s_1|^{4/\epsilon})
 N^{-1+\epsilon}.
\label{main5}\end{equation}

To compute the averages of the type $\la \ti R_{l,l'}\ra$ we use
another tool. Denote $\D_l=\frac{d}{dS_1^{(l)}}$. One can see
easily that, e.g., $E\{\la \ti R_{1,2}\ra\}$ can be represented in
the form
$$
E\{\la \ti R_{1,2}\ra\}=E\bigg\{\frac{\la \D_1\D_2
G(S_1^{(1)})G(S_1^{(2)})\ra_{(p-1)}}
{\la G(S_1^{(1)})G(S_1^{(2)})\ra_{(p-1)}}\bigg\},
$$
where $G(S)=e^{g(S)}$ and the symbol $\la...\ra_{(p-1)}$ means the averaging
with respect to the Hamiltonian (\ref{conc}) in which $g(S_1)$ is replaced by
$0$.

Let us again consider a standard Gaussian variable
$u$ and introduce a function
\begin{equation}
G_t(S,u)=
\frac{1}{\sqrt{(1-t)(R-q+\e)}}\int e^{-x^2/2(R-q+\e)(1-t)}
G(\sqrt t S+u\sqrt{q(1-t)}+x)dx .
\label{G_t}\end{equation}
In particular,
\begin{equation}
G_0(u)= \frac{1}{\sqrt{(R-q+\e)}}
\int e^{-x^2/2(R-q+\e)} G(u\sqrt{q}+x)dx.
\label{G_0}\end{equation}
 It is evident, that $G_0(S,u)=G_0(u)$ (i.e. it does not depend on $S$) and
 $G_1(S,u)=G(S)$. We remark, that
the definition (\ref{G_t}) becomes more natural, if we introduce it through
the Fourier transform $\hat G(\lambda)$ of the function $G(S)$:
\begin{equation}
G_t(S,u)=
\frac{1}{\sqrt{2\pi}}\int\hat G(\lambda)\exp\{-i\lambda(S\sqrt t+u\sqrt{q(1-t)})-
\frac{1-t}{2}(R-q+\e)\lambda^2\}.
\label{Four}\end{equation}
Now for any $\xi^{(1)}_i$-independent function
$f:{\bf R}^{N\times n}\to{\bf R}$  and
some polynomial $P(x_1,\dots,x_n)$ consider the operator
$\PP_t=P(t^{-1/2}\D_1,...,t^{-1/2}\D_n)$
\begin{equation}\begin{array}{lll}
\varphi_t^{(n)}(f\PP_t)&=&\ds{E\bigg\{\frac{\la f\,\PP_t
G_t(S_1^{(1)},u)...G(S_1^{(n)},u)\ra_{(p-1)}}
{\la G_t(S_1^{(1)},u)...G_t(S_1^{(n)},u)\ra_{(p-1)}}\bigg\}}\\
&=&\ds{E\left\{\lla f\,\frac{\PP_t
G_t(S_1^{(1)},u)...G(S_1^{(n)},u)}{G_t(S_1^{(1)},u)...G(S_1^{(n)},u)}
\rra_{(t)}\right\},}
\end{array}\label{phi_t}\end{equation}
where $\la...\ra_{(t)}$ means  the Gibbs averaging corresponding to the
$n$ replicas of the Hamiltonian (\ref{conc}) in which $g(S_{1})$ is
substituted by $\log G_t(S_1,u)$. According to the result of \cite{BL},
this function is also concave with respect to $S_1$ and so inequalities
(\ref{main1}) and (\ref{main2}) for it are also valid.

We remark here also that due to the definition $G_t$ (\ref{G_t}) the operator
$\PP_t$  has a natural form:
\begin{equation}\begin{array}{lll}
\PP_t G_t(S_1^{(1)},u)...G(S_1^{(n)},u)
&=&\ds{\frac{1}{(\sqrt{2\pi})^n}\int\hat G(\lambda_1)\dots
\hat G(\lambda_n)P(-i\lambda_1,\dots,-i\lambda_n)}\\
&&\ds{
\exp\{-i\sum\lambda_lS_1^{(l)}\sqrt t-iu\sum\lambda_l\sqrt{q(1-t)}-
\frac{1-t}{2}(R-q)\lambda^2\}.}
\end{array}\label{Four2}\end{equation}
So for $t=0$ it is well defined:
$$
\PP_t G_t(S_1^{(1)},u)...G(S_1^{(n)},u)\bigg|_{t=0}=P(q^{-1/2}\frac{d}{dx_1},
\dots,
q^{-1/2}\frac{d}{dx_1})G_0(x_1)...G_0(x_n)\bigg|_{x_1=...=x_n=u}.
$$

 Let us compute the derivative with respect to $t$ of $\varphi_t^{(n)}(f\PP_t)$.
\begin{equation}\begin{array}{lll}\ds{
\frac{d}{dt}\varphi_t^{(n)}(f\PP_t)}&=&\ds{\frac{1}{2}\sum_{l=1}^n
\varphi_t^{(n)}(f(R_{l,l}-R)t^{-1}\D_l^2\PP_t)
-\frac{n}{2}\varphi_t^{(n+1)}(f(R_{n+1,n+1}-R)t^{-1}\D_{n+1}^2\PP_t)}
\\
&&\ds{
+\sum_{l<l'}^n \varphi_t^{(n)}(f(R_{l,l'}-q)t^{-1}\D_l\D_{l'}\PP_t)
-n\sum_{l=1}^n\varphi_t^{(n)}(f(R_{l,l'}-q)t^{-1}\D_l\D_{n+1}\PP_t) }\\
&&\ds{
+\frac{n(n+1)}{2}\varphi_t^{(n+2)}(f(R_{n+1,n+2}-q)t^{-1}\D_{n+1}
\D_{n+2}\PP_t)}.
\end{array}\label{phi'}\end{equation}
This formula can be obtained easily if we  differentiate with respect to $t$ and then
integrate by parts with respect to $\xi_i^{(1)}$ and $u$
in the expressions (\ref{Four}) and (\ref{Four2}).
\begin{proposition}\label{pro:6} For any polynomial
$P(\l_1,\dots\l_n)$ ($\emph{deg}P\le 6$)
\begin{equation}
|\varphi_t^{(n)}(\PP_t)|\le C,
\label{p6.1}\end{equation}
where constant $C$ depends on $n$ and on the polynomial $P(x_1,...,x_n)$.
\end{proposition}
\begin{remark} If we take $g(x)=-\log\emph{H}(\frac{x}{\sqrt\e})$,
then Proposition \ref{pro:6} is valid for the polynomial of any
degree.
\end{remark}
As it was mentioned above
 the inequalities (\ref{bl.1}) and (\ref{bl.2})
for $\la...\ra_{(t)}$ are also valid. Therefore the estimate (\ref{main1}), and
(\ref{main3}) are fulfilled and so, using the Schwartz inequality,
Proposition \ref{pro:6} and (\ref{main3}), we obtain:
\begin{equation}
|\varphi_1^{(n)}(f\PP_1)-\varphi_0^{(n)}(f\PP_0)
|\le C\max_t\nu_t^{1/4}(|f|^4) N^{-1/2}.
\label{main6}\end{equation}
Using the same formula to compute the second derivative of
$\varphi_t^{(n)}(f\PP_t)$
with respect to $t$, we obtain the expression in each term of which
we have $( R_{l,l'}-q)( R_{l_1,l'_1}-q)$ or $(\ti R_{l,l'}-q)(R_{l_1,l_1}-R)$
or $(R_{l_1,l_1}-R)^2$.  Using the  H\"{o}lder  inequality,
Proposition \ref{pro:6} and (\ref{main3}), we obtain:
\begin{equation}
|\varphi_1^{(n)}(f\PP_1)-\varphi_0^{(n)}(f\PP_0)
-\frac{d}{dt}\varphi_0^{(n)}(f\PP_0)|
\le C\max_t\nu_t^{1/4}(|f|^4) N^{-1}.
\label{main7}\end{equation}

\medskip

\no{\it Proof of Theorem \ref{thm:2}}

\no We prove Theorem \ref{thm:2} in 3 steps which are Lemma \ref{lem:3},
\ref{lem:4} and \ref{lem:5}.
 \begin{lemma}\label{lem:3}
Consider an expression of the form $T_{1,2}^kP$ where $P$
is some product of the terms $T_{i,j}$ (different from
 $T_{1,2}$), $T_i$ and $\dq$. Then
\begin{equation}
E\la T_{1,2}^kP\ra=(k-1)AE\la T_{1,2}^{k-2}P\ra+
O(N^{-1/2+\epsilon}),
\label{l3.1}\end{equation}
where
\begin{equation}
A=\frac{b_0}{1-\alpha b_0c_0},
\label{A}\end{equation}
with
\begin{equation}
b_0=(R-q)^{2},\quad c_0=q^{-2}E\{ (g_0''(u))^2\},\quad g_0(u)=\log G_0(u),
\label{b_0,c_0}\end{equation}
where $E\{..\}$  means the averaging with respect to the standard
Gaussian random variable $u$ and $G_0(u)$ was defined in (\ref{G_0}).

\end{lemma}
\begin{lemma}\label{lem:4}
Consider an expression of the form $T_{1}^kP$ where $P$
is some product of the terms  $T_i$ with $i\not=1$ and $\dq$. Then
\begin{equation}
E\la T_{1}^kP\ra=(k-1)BE\la T_{1}^{k-2}P\ra+O(N^{-1/2+\epsilon})
\label{l4.1}\end{equation}
where $B$ is some $N$-independent constant which is an algebraic
expression of the coefficients $b_0$, $c_0$ and
\begin{equation}\begin{array}{l}
b_1=E\{\la \dot s^2\ra_0\la s\ra_0^2\}=q(R-q)\\
c_1=E\{\la(\D_1-\D_2)\D_1\D_3\D_4 \ra_{(0)}\}=q^{-2}E\{g_0''(g_0')^3\}\\
c_2=E\{\la (\D_1-\D_2)\D_1^2\D_3\ra_{(0)}\}=q^{-2}E\{(g_0'''+2g''_0g_0')g_0'\}\\
c_3=E\{\la (\D_1^2-\D_2^2)\D_1^2\ra_{(0)}\}=q^{-2}E\{(g_0^{(4)}+4g_0'''g_0'+
2(g''_0)^2+4g''_0(g_0')^2)\}\\
c_4=E\{\la\D_1\D_2\D_3\D_4\ra_{(0)}\}=q^{-2}E\{(g_0')^4\}.
\end{array}\label{b,c}\end{equation}
For $h=0$
\begin{equation}
B=2\frac{\d^2F}{\d z^2}+\frac{1}{2}A
\label{B}\end{equation}
\end{lemma}
\begin{lemma}\label{lem:5}
\begin{equation}
E\la \dq ^k\ra= (k-1)CE\la \dq ^{k-2}\ra,
\label{l5.1}\end{equation}
where $C$ is some $N$-independent constant which is an algebraic
expression of the coefficients $b_{0,1}$, $c_{0,1,2,3,4}$.
\end{lemma}
One can see easily that the statement of Theorem \ref{thm:2}
follows from these lemmas by induction.
So our goal is to prove  the Lemmas.

\smallskip

\no {\it Proof of Lemma \ref{lem:3}}.
First of all we rewrite all terms of our initial product (including
all $T_{1,2}$) in the form
\begin{equation}\begin{array}{l}
T_{l,l'}=N^{-1/2}(\bJ^{-(l)}-\bJ^{-(k)})(\bJ^{-(l)}-\bJ^{-(k')})+
N^{-1/2}(J^{(l)}_1-J^{(k)}_1)(J^{(l)}_1-J^{(k')}_1)\\
T_{l}=N^{-1/2}(\bJ^{-(l)}-\bJ^{-(k_1)})\bJ^{-(k_1')}+
N^{-1/2}(J^{(l)}_1-J^{(k_1)}_1)J^{(k_1')}_1\\
\dq=N^{-1/2}((\bJ^{-(k_2)},\bJ^{-(k_2')})-(N-1)q)
+N^{-1/2}(J^{(k_2)}_1J^{(k_2')}_1-q),
\end{array}\label{l3.2}\end{equation}
where indexes $k,k',k_1,k_1',k_2,k_2'$ are different for each
term in the product and different from initial indexes $l,l'$.
We denote the last term in  $i$-th expression by
$N^{-1/2}f_{i}(J_1)$.
 Using the symmetry of the Hamiltonian and  the above
 representation,  we can write
 \begin{equation}\begin{array}{lll}
 E\{T_{1,2}^kP\}&=&
 \sqrt N\nu_1((s_1-s_k)(s_2-s_{k'}) \ti P^{-})\\
&& \ds{
 +  \sum_i\nu_1((s_1-s_k)(s_2-s_{k'})f_i(s) \ti P^{-}_i)
 +O(N^{-1/2})}\\
 &=&I+II+O(N^{-1/2}) ,
\end{array} \label{l3.3}\end{equation}
where $\ti P^{-}$ means the product only of such terms of (\ref{l3.2})
which does not contain
$s_l$ (including that, corresponding to $T_{1,2}$)
 and $\ti P^{-}_i$ means the product of the same terms except the
$i$-th one. The term $O(N^{-1/2})$ appears because of the products which contain
more than 1 term $f_i(s)$. Applying  formula
(\ref{main4}) to the term $II$,  we have got
$$\begin{array}{lll}
II&=&\ds{\sum_i\nu_0((s_1-s_k)(s_2-s_{k'})f_i(s))
\nu_1( \ti P^{-}_i)+O(N^{-1/2})}\\
&=&\ds{
\sum_i\nu_0((s_1-s_k)(s_2-s_{k'})f_i(s))\nu_1(\ti P_i)
+O(N^{-1/2})}.
\end{array}$$
But, if $f_i(s)$ does not contain both $s_1$ and $s_2$,
$$
\nu_0((s_1-s_k)(s_2-s_{k'})f_i(s))=0.
$$
So we obtain
\begin{equation}
II=(k-1)b_0E\{\la T_{1,2}^{k-2}P\ra\}+O(N^{-1/2}).
\label{l3.4}\end{equation}
Now let us analyze term $I$, using formula (\ref{main5}).
It is evident that $\nu_0$ term here is equal to $0$. Calculating
$\nu'_0$, we get
$$
I=\sqrt N\sum_{l<l'}^n\nu_0((s_1-s_k)(s_2-s_{k'})s_ls_{l'})
\nu_1(\ti R_{l,l'}^{-}-d)\ti P^{-})+O(N^{-1/2+\epsilon}).
$$
All the rest terms in (\ref{nu'}) disappear because
$$
\nu_0((s_1-s_k)(s_2-s_{k'})s_ls_{l'})=b_0(\delta_{l,1}\delta_{l',2}+
\delta_{l,k}\delta_{l',k'}-\delta_{l,k}\delta_{l',2}-
\delta_{l,k'}\delta_{l',1}).
$$
So, we have got
\begin{equation}\begin{array}{lll}
I&=&b_0\sqrt N
E\{\la(\ti R_{1,2}-\ti R_{1,k'}-\ti R_{2,k}+\ti R_{k,k'})\ti P\ra\}
+O(N^{-1/2+\epsilon})\\
&=&b_0III+O(N^{-1/2+\epsilon}).
\end{array}\label{l3.5}\end{equation}
To analyze  $III$ we use again the symmetry of (\ref{conc}) and
notations of (\ref{phi_t}) to write
$$\begin{array}{lll}
\alpha^{-1}III&=&\sqrt NE\{\la (g'(S^{(1)}_1)-g'(S^{(k)}_1))(g'(S^{(2)}_1)-g'(S^{(k')}_1))
\ti P\ra\}\\
&=&N^{1/2}\varphi_1((\D_1-\D_k)(\D_2-\D_{k'})\ti P).
\end{array}$$
Now, applying formula (\ref{main7}), we can write
\begin{equation}
\alpha^{-1}III=\sqrt N\sum_{l<l'}\phi_0((\D_1-\D_k)(\D_2-\D_{k'})\D_l\D_{l'})
E\{\la(R_{l,l'}-q)\ti P\ra\}+O(N^{-1/2+\epsilon}).
\label{l3.6}\end{equation}
All the rest terms in (\ref{phi'}) disappear because
$$
\varphi_0((\D_1-\D_k)(\D_2-\D_{k'}))=0,
\quad \varphi_0((\D_1-\D_k)(\D_2-\D_{k'})\D_l^2)=0.
$$
Let us remark that
$$
\varphi_0((\D_1-\D_k)(\D_2-\D_{k'})\D_l\D_{l'})=c_0(\delta_{l,1}\delta_{l',2}+
\delta_{l,k}\delta_{l',k'}-\delta_{l,k}\delta_{l',2}-
\delta_{l,k'}\delta_{l',1}),
$$
with $c_0$ defined in (\ref{b_0,c_0}).

Hence we get from (\ref{l3.6}) that
\begin{equation}\begin{array}{lll}
\alpha^{-1}III&=&c_0\sqrt NE\{\la( R_{1,2}- R_{1,k'}- R_{2,k}+ R_{k,k'})\ti P\ra\}
+O(N^{-1/2+\epsilon})\\
&=&c_0E\{\la T_{1,2}^kP\ra\}+O(N^{-1/2+\epsilon}).
\end{array}\label{l3.7}\end{equation}
Now, using relations (\ref{l3.2}), (\ref{l3.3}), (\ref{l3.4}),
(\ref{l3.5}), (\ref{l3.6}) and (\ref{l3.7}), we get (\ref{l3.1}).
Lemma \ref{lem:3} is proven.

\medskip

\no{\it Proof of Lemma \ref{lem:4}}

\no Like in the proof of Lemma \ref{lem:2} we use representation
(\ref{l3.2}) for all the terms and let for the first $T_1$ the
numbers of replicas here are $1,2,3$ and for the $i$-th $T_1$
they are $(1,l_i,l_{i}+1)$.  Using the symmetry
of the problem similarly to (\ref{l3.3}) write
\begin{equation}\begin{array}{lll}
E\{T_{1}^kP\}&=&
 \sqrt N\nu_1((s_1-s_2)s_3 \ti P^{-})\\
&& \ds{
 +  \sum_{i=2}^k\nu_1((s_1-s_2)s_3(s_1-s_{k_i})s_{k'_i}f_i(s) \ti P^{-}_i)
 +O(N^{-1/2})}\\
 &=&I+II+O(N^{-1/2}) ,
\end{array}\label{l4.2}\end{equation}
where $\ti P^{-}$ means the product only of such terms of (\ref{l3.2})
which does not contain
$s_1$  and $\ti P^{-}_i$ means the product of the same terms except the
$i$-th one. By the same way as in Lemma \ref{lem:2} we get
\begin{equation}
II=(k-1)b_1E\{T_{1}^{k-2}P\}+ O(N^{-1/2}),
\label{l4.3}\end{equation}
where $b_1$is defined by (\ref{b,c}).
Using the formula (\ref{main5}) we have got
\begin{equation}\begin{array}{lll}
\alpha^{-1}I&=&\ds{\sqrt N\bigg[2b_1E\{\la(\ti U_1-\ti U_2)\ti P\ra\}
+b_0E\{\la(\ti R_{1,3}-\ti R_{2,3})\ti P\ra}\\
&&\ds{
+b_1\sum_{l=3}^n E\{\la(\ti R_{1,l}-\ti R_{2,l})\ti P\ra\}
-nb_1E\{\la(\ti R_{1,n+1}-\ti R_{2,n+1})\ti P\ra\}\bigg]
   +O(N^{-1/2+\epsilon})}.
\end{array}\label{l4.4}\end{equation}
Using again the symmetry and notations (\ref{phi_t}), we write
$$\begin{array}{lll}
\alpha^{-1}I&=&\ds{\sqrt N\bigg[2b_1\varphi_1((\D_1^2-\D_2^2)\ti P)+
b_0\varphi_1((\D_1-\D_2)\D_3\ti P)}\\
&&\ds{
+b_1\sum_{l=3}^n\varphi_1((\D_1-\D_2)\D_l\ti P)
-nb_1\varphi_1(\D_1-\D_2)\D_{n+1}\ti P)\bigg]
   +O(N^{-1/2+\epsilon})}.
\end{array}$$
Applying formula (\ref{phi'}) we get
\begin{equation}\begin{array}{lll}
\alpha^{-1}I&=&\ds{\sqrt N\bigg[a^{(1)}_1E\{\la(R_{1,1}-R_{2,2})\ti P\ra\}
+a^{(1)}_2E\{\la( R_{1,3}- R_{2,3})\ti P\ra\}}\\
&&\ds{
+a^{(1)}_3\sum_{l=3}^n
E\{\la( R_{1,l}- R_{2,l}-R_{1,n+1}+ R_{2,n+1}))\ti P\ra\}\bigg]
   +O(N^{-1/2+\epsilon})}\\
& =&a^{(1)}_1 E\{\la(R_{1,1}-R_{2,2})\ti P\ra\}\sqrt N
+a^{(1)}_2 E\{T_1^kP\}\\
&&  +(k-1)Aa^{(1)}_3E\{T_1^{k-2}P\}
 +O(N^{-1/2+\epsilon}),
\end{array}\label{l4.5}\end{equation}
where $a^{(1)}_{1,2,3}$ are  some algebraic combinations of
$b_{0,1}$ and $c_{0,1,2,3,4}$ defined by (\ref{b,c}).

Here we have used Lemma \ref{lem:3}, according to which
$$
\sqrt NE\{\la( R_{1,l}- R_{2,l}-R_{1,n+1}+ R_{2,n+1})\ti P\ra\}=
E\{\la T_{1,l}^2T^{k-2}P\ra\}=AE\{T_1^{k-2}P\}+O(N^{-1/2+\epsilon})
$$
for $l=l_i$ $(i=2\dots,k)$ and it is zero for the rest of $l$.
Now we are faced with a problem to compute
$E\{\la(R_{1,1}-R_{2,2})\ti P\ra\}\sqrt N$.

By using the same procedure we can get
\begin{equation}\begin{array}{lll}
\sqrt NE\{\la(R_{1,1}-R_{2,2})T_1^{k-1} P\ra\}&=&2b_1(k-1)E\{T_1^{k-2}P\}\\
&&+\alpha a^{(2)}_1E\{\la(R_{1,1}-R_{2,2})T_1^{k-1} P\ra\}\sqrt N\\
&&+\alpha a^{(2)}_2E\{T_1^kP\}
+(k-1)\alpha a^{(2)}_3E\{T_1^{k-2}P\}+O(N^{-1/2+\epsilon}),
\end{array}\label{l4.6}\end{equation}
where $a^{(2)}_{1,2,3}$ are  some algebraic combinations of
$b_{0,1}$ and $c_{0,1,2,3,4}$ defined by (\ref{b,c}).

Now we have got the system of two equations with respect to $E\{T_1^kP\}$
and

$E\{\la(R_{1,1}-R_{2,2})T_1^{k-1} P\ra\}\sqrt N$. This system  gives us
\begin{equation}
E\{T_1^kP\} =(k-1)BE\{T_1^{k-2}P\},\quad
E\{\la(R_{1,1}-R_{2,2})T_1^{k-1} P\ra\}\sqrt N=(k-1)\ti BE\{T_1^{k-2}P\}\sqrt N
\label{l4.7}\end{equation}
with some $B$ and $\ti B$. But using the fact that $B$ and $\ti B$ do not depend
on $k$, we observe that
$$\begin{array}{lll}\ds{
2\frac{\d^2F}{\d z\d h}}&=&\ds{N^{-1}\sum_{i,j}E\{\la J_i^2\dot J_j\ra \ti h_j\}=
4\frac{\d^2F}{\d z^2}-2E\{\la R_{1,1}T_1\ra\}}\\
&=&\ds{
4\frac{\d^2F}{\d z^2}-2\ti B}.
\end{array}$$
Then since  $\ds{\frac{\d^2F}{\d z\d h}=0}$ for $h=0$, we get
\begin{equation}
\ti B=2\frac{\d^2F}{\d z^2}.
\label{ti_B}\end{equation}
Similarly one can get
$$\begin{array}{lll}\ds{
\frac{\d^2F}{\d h^2}}&=&\ds{N^{-1}\sum_{i,j}E\{\ti h_i\la \dot J_i\dot J_j\ra \ti h_j\}
=R-q+2\frac{\d^2F}{\d z\d h}
-4\ti B+4E\{\la T_1^2\ra\}-2E\{\la T_{1,2}^2\ra\}}\\
&=&\ds{
R-q-4\ti B+4B-2A.}
\end{array}$$
Thus, using the fact that $\ds{\frac{\d^2F}{\d h^2}=R-q}$ for $h=0$,
we have got formula (\ref{B}) for $B$.

\medskip

\no{\it Proof of Lemma \ref{lem:5}}

We write
\begin{equation}\begin{array}{lll}
E\{\dq^k\}&=&
 \sqrt N\nu_1((s_1s_2-q)\ti P^{-})\\
&& \ds{
 +  \sum_{i=2}^k\nu_1((s_1s_2-q)^2 \ti P^{-}_i)
 +O(N^{-1/2})}\\
 &=&I+II+O(N^{-1/2}) ,
\end{array}\label{l5.2}\end{equation}
where
$$
\ti P^{-}=\prod_{l=2}^kN^{1/2}(R_{2l-1,2l}-q)
$$
One can see easily
\begin{equation}
II=(k-1)q^2E\{\dq^{k-2}\}+O(N^{-1/2})
\label{l5.3}\end{equation}
Calculating $I$ with (\ref{main4}) we get
\begin{equation}\begin{array}{lll}
I&=&\sqrt N\bigg[
 q^2\sum_{l<l'}\nu_1((\ti R_{l,l'}-\ti R_{l,n+1}-\ti R_{l',n+2}+
 \ti R_{n+1,n+2})\ti P^{-})\\
&& \ds{
 -(2b_1-q^2) \sum_{l\ge 3}\nu_1((\ti R_{1,2}-\ti R_{l,2})\ti P^{-})
-\frac{q^2}{2}\sum_{l\ge 3}\nu_1((\ti R_{1,1}-\ti R_{l,l})\ti P^{-}) }\\
 && \ds{
+(b_0-2b_1)\nu_1((\ti R_{1,2}-d)\ti P^{-})
+2b_1\nu_1((\ti R_{1,1}-U)\ti P^{-})\bigg]+O(N^{-1/2+\epsilon}).}
\end{array}\label{l5.4}\end{equation}
Here we have used that since $\ti P$ does not contain replicas $1$ and $2$
we can replace $n+1,n+2\to 1,2$
Now we apply formula (\ref{main7})
 using the following relations:
$$\begin{array}{rll}
E\{\la T_{l,l'}\ti P\ra\}&=&\ds{A\sum_{k=2}\delta_{l,2k-1}\delta_{l',2k}
E\{\dq^{k-2}\}\,\, (l<l'),}\\
E\{\la T_{l}\ti P\ra\}&=&BE\{\dq^{k-2}\},\\
\sqrt NE\{\la(R_{1,1}-R_{l,l})\ti P\ra\}&=&-\ti BE\{\dq^{k-2}\}.
\end{array}$$
These relations follows from Lemmas \ref{lem:3}, \ref{lem:4} and
formula (\ref{ti_B}).
\begin{equation}\begin{array}{lll}
\alpha^{-1}I&=&\ds{a^{(3)}_1 E\{\la(R_{1,1}-R)\dq^{k-2}\ra\}\sqrt N}\\
&&+a^{(3)}_2 E\{\dq^{k}\}
 +(k-1)a^{(3)}_3E\{\dq^{k-2}\}
 +O(N^{-1/2+\epsilon}).
\end{array}\label{l5.5}\end{equation}
So we have got the equation
\begin{equation}\begin{array}{lll}
 E\{\dq^{k}\}&=&\ds{(k-1)(q^2+\alpha a^{(3)}_3)E\{\dq^{k-2}\} }\\
 &&+ \alpha a^{(3)}_1 E\{\la(R_{1,1}-R)\dq^{k-2}\ra\}\sqrt N
+\alpha a^{(3)}_2 E\{\dq^{k}\}
  +O(N^{-1/2+\epsilon}),
\end{array}\label{l5.6}\end{equation}
where $a^{(3)}_{1,2,3}$ are  some algebraic combinations of
$b_{0,1}$ and $c_{0,1,2,3,4}$ defined by (\ref{b,c}).

By the same way, studying $\sqrt NE\{\la(R_{1,1}-R)\ra\dq^{k-1}\}$, we get
the equation
\begin{equation}\begin{array}{lll}
 \sqrt NE\{\la(R_{1,1}-R)\ra\dq^{k-1}\}&=&\ds{(k-1)(q^2
 +\alpha a^{(4)}_3)E\{\dq^{k-2}\} }\\
&& + \alpha a^{(4)}_1 E\{\la(R_{1,1}-R)\ra\dq^{k-2}\}\sqrt N\\
&& +\alpha a^{(4)}_2 E\{\dq^{k}\}
  +O(N^{-1/2+\epsilon}),
\end{array}\label{l5.7}\end{equation}
where $a^{(4)}_{1,2,3}$ are  some algebraic combinations of
$b_{0,1}$ and $c_{0,1,2,3,4}$ defined by (\ref{b,c}).

Considering (\ref{l5.6}) and (\ref{l5.7}) as a system of equations with respect
to $ E\{\dq^{k}\}$ and

$\sqrt N E\{\la(R_{1,1}-R)\ra\dq^{k-1}\}$,
 we finish the proof of Lemma~\ref{lem:5}.

\section{Auxiliary results}

{\it Proof of Proposition \ref{pro:2}}
Consider any ${\bf x},{\bf y}\in {\bf R}^N$ and  write
$$
(\mathcal{A}^{(f)}{\bf x},{\bf y})=
\la (\dot{\bJ},{\bf x})(\dot{\bJ},{\bf y})f\ra\le
\la (\dot{\bJ},{\bf x})^2(\dot{\bJ},{\bf y})^2\ra^{1/2}
\la f^2\ra^{1/2}\le
\frac{3\la f^2\ra^{1/2}}{z}|{\bf x}|\,|{\bf y}|,
$$
where we have used inequality (\ref{bl.1}).
Similarly, using inequalities (\ref{bl.1}) and (\ref{bl.2}),
 for any ${\bf x}\in{\bf R}^p,{\bf y}\in {\bf R}^N$
$$\begin{array}{lll}
(\mathcal{B}^{(f)}{\bf x},{\bf y})&=&
\ds{\la (\dot{\bJ},{\bf x})^4\ra^{1/4}\la f^4\ra^{1/4}
\lla\sum_\mu (g'(S_\mu)-\la g'(S_\mu)\ra)^2y_\mu^2\rra^{1/2}}\\
&\le&
\ds{\frac{\la 3 f^4\ra^{1/4}||\mathcal{A}_*||^{1/2}\la|g''|^2\ra^{1/2}}{z}
|{\bf x}|\,|{\bf y}|}.
\end{array}$$
The inequality for the matrix $\mathcal{C}$ can be proven by the
same way.

\smallskip
\no {\it Proof of Proposition \ref{pro:3}}

\no From (\ref{cond_g}) one can easily derive that
$$
(g'(S))^2\le C_1-C_2 g(S).
$$
Thus it is enough to prove that
$$
\lla -\frac{1}{N}\sum g(S_\mu)\rra\le C.
$$
Define the Hamiltonian
$$
-\ti\H(\tau)=\tau\sum g(S_\mu)+z(\bJ,\bJ).
$$
Let $\la...\ra_\tau$ be a corresponding Gibbs average. One can see easily
that
$$
\varphi(\tau)\equiv -\lla \frac{1}{N}\sum g(S_\mu)\rra.
$$
is a decreasing  function of $\tau$. Thus
$$\begin{array}{lll}\ds{
\lla -\frac{1}{N}\sum g(S_\mu)\rra}&=&\ds{\varphi(1)\le\varphi(0)=
\lla -\frac{1}{N}\sum g(S_\mu)\rra_0}\\
&\le&\ds{
 \frac{C}{N}\la\sum
\la S_{\mu}^2\ra_0\le \frac{C}{zN^2}\sum(\bxm,\bxm)}.
\end{array}$$

\medskip

\no {\it Proof of Lemma \ref{lem:2}}

\no We prove first that
\begin{equation}\begin{array}{c}
|\nu_t(\ti R_{1,2})-\nu_0(\ti R_{1,2})|\le \frac{C}{\sqrt N},\quad
|\nu_t(\ti U_1)-\nu_0(\ti U_1)|\le \frac{C}{\sqrt N},\\
|\varphi_t( R_{1,2})-\varphi_0( R_{1,2})|\le \frac{C}{\sqrt N},\quad
|\varphi_t( R_{1,1})-\varphi_0( R_{1,1})|\le \frac{C}{\sqrt N}.
\end{array}\label{l2.2}\end{equation}
 To this end consider the Hamiltonian $H_N(t)$ which has the form
 (\ref{H(t)}) with
 $d,U$ substituted by
 $$d_N=\nu_0(g'(S_\mu^{(1)})
 g'(S_\mu^{(2)})),\quad U_N=
\nu_0( g''(S_\mu)+g'^2(S_\mu))
$$
respectively.
Then we use formula (\ref{nu'}) for $f=\sqrt N \ti R_{1,2}$ and
$f=\sqrt N \ti U_{1}$, but we
write it in the form:
\begin{equation}\begin{array}{lll}
\nu_t'(f)&=&\ds{\frac{1}{2}\sum_{l=1}^n \nu_t((f-\la f\ra_t) s_l^2( U_l^{-}-U_N))
+\sum_{l<l'}^n\nu_t((f-\la f\ra_t)s_ls_{l'}
(\ti R_{l,l'}^{-}-d_N))}\\
&&\ds{
-n\sum_{l=1}^n\nu_t((f-\la f\ra_t)s_ls_{n+1}(\ti R_{l,n+1}^{-}-d_N))}\\
&&\ds{+\frac{n(n+1)}{2}\nu_t((f-\la f\ra_t)s_{n+1}s_{n+2}
(\ti R_{n+1,n+2}^{-}-d_N))}.
\end{array}\label{l2.3}\end{equation}
Using the Schwartz inequality and (\ref{bl.2}), due to
the terms $\nu_t((f-\la f\ra_t)^2)$ we obtain the first line of
(\ref{l2.2}).
But then, on the basis of (\ref{l2.2}) one can derive from (\ref{l2.3})
that the first line of (\ref{l2.2}) is valid even if we replace
$C N^{-1/2}$ by $C N^{-1}$. Now
similarly to (\ref{main3}) one can conclude that
for any $r>2$
\begin{equation}
E\{\la|\ti R_{l,l'}-d_N|^r\}\le\frac{C}{N},\quad
E\{\la|\ti U_l-U_N|^r\ra\}\le\frac{C}{N}.
\label{l2.4}\end{equation}
Similarly, using (\ref{phi'}) for $f=\sqrt N R_{1,2}$ and
$ f=\sqrt N R_{1,1}$ with $q_N=\phi_0(s_1s_2)$,
$R_N=\phi_0(s_1^2)$, we prove first the second line of
(\ref{l2.2}). Then,  by the same way as above, we get
\begin{equation}
E\{\la| R_{l,l'}-q_N|^{2}\ra\}\le\frac{C}{N},\quad
E\{\la|R_{l,l}-R_N|^2\ra\}\le\frac{C}{N}.
\label{l2.5}\end{equation}
Now we remark that since it was proved
in \cite{ST3}, that the system (\ref{q,R}) has a unique solution,
to prove  (\ref{l2.1}) it is enough to show that our $q_N$,
$R_N$ satisfy this system with the  error terms
$O(N^{-1+\epsilon})$ and $d_N$, $U_N$ satisfy relations (\ref{d,U})
with the same error.
Now, on the basis of (\ref{l2.4}) and (\ref{l2.5})
 it can be shown  easily by
 formulas (\ref{nu'}), (\ref{main5}) with
$f(s)=s_1s_2$ and $f(s)=s_1^2$ and by formulas
(\ref{phi'}) and (\ref{main7}) with $f=\D_1\D_2$ and
$f=\D_1^2$.

\medskip

\no {\it Proof of Proposition \ref{pro:4}}

\no Let us  denote
\begin{equation}\begin{array}{l}\ds{
\phi (s)=\log\int d\bJ^{-}\exp\{-H_t(\bJ^{-},s)\}=
\phi_t(s)+us\sqrt{d(1-t)}-\frac{s^2}{2}(z-(1-t)(U-d))}\\
\ds{
\Rightarrow
\la s^n\ra_t=\la s^n\ra_\phi=\frac{\int s^ne^{\phi (s)}ds}
{\int e^{\phi (s)}ds}}.
\end{array}\label{p4.2}\end{equation}
According to the results \cite{BL}, $\phi_t(s)$ is a concave
function. Besides, there exists $\delta>0$ such that
$$(z-(1-t)(U-d))\ge (z-(U-d))=(R-q)^{-1}>\delta.
$$
Then, according to the results \cite{BL},
\begin{equation}
\la |s-\la s\ra_\phi|^n\ra_\phi\le \delta^{-n}2^n\Gamma(n).
\label{p4.3}\end{equation}
So, to prove  Proposition \ref{pro:4} it is
enough to estimate $\la s\ra_\phi$.

Denote $s^*$ the point of maximum of the function
$\phi (s)$. Then it follows from representation (\ref{p4.2})
of the function $\phi (s)$ that
$$
|s^*|\le \delta^{-1}|\varphi_t'(0)+u\sqrt{d(1-t)}|.
$$
On the other hand, by \cite{BL}
$$
\la |s-s^*|\ra_\phi\le \delta^{-1}\quad\Rightarrow
\quad|\la s\ra_\phi|\le \delta^{-1}(1+|\phi_t'(0)+\sqrt{d(1-t)}u|)
$$
Now, since
$$
\phi_t'(0)=\sum_\mu\frac{\xmo }{\sqrt N}\la g'(S^{-}_\mu)\ra_0
$$
and $\la...\ra_0$ does not depend on $\xmo$, we have
$$
E\{\la s\ra^n\}\le 2^n\delta^{-n}\Gamma(n)\bigg(d^n+E\bigg\{\bigg(
N^{-1}\sum_\mu\la g'(S^{-}_\mu)\ra_0^2\bigg)^{n/2}\bigg\}\bigg).
$$
Then using Proposition \ref{pro:3}, we obtain the statement
of Proposition \ref{pro:4}.

\medskip

\no {\it Proof of Proposition \ref{pro:6}}

\no According to the representation (\ref{Four2}), $\PP_t$ is some
polynomial of the derivatives $\frac{\partial^k}{\partial
S^k}\log G_t(S_1^{(j)},u)$ ($k=1,..,6$, $j=1,...,n$). But under condition
(\ref{cond_g}) for $k\ge 2$ these derivatives are uniformly
bounded functions. So we need only to prove that
\begin{equation}
E\left\{\lla\bigg(\frac{\partial}{\partial
S}\log G_t(S_1^{(j)},u)\bigg)^{2k}\rra_{(t)}\right\}\le C(k).
\label{p6.2}\end{equation}
Similarly to the proof of Proposition \ref{pro:4} by
(\ref{bl.2}) and (\ref{cond_g}) inequality (\ref{p6.2}) can be
derived from the inequality
\begin{equation}
E\left\{\lla\bigg(\frac{\partial}{\partial
S}\log G_t(S_1^{(j)},u)\bigg)^2\rra_{(t)}^{k}\right\}\le C(k).
\label{p6.3}\end{equation}
But similarly to the proof of Proposition \ref{pro:3}, since
$$
\bigg(\frac{\partial}{\partial
S}\log G_t(S_1^{(j)},u)\bigg)^2\le C_1-C_2 \log G_t(S_1^{(j)},u),
$$
one can get that
$$
\lla\bigg(\frac{\partial}{\partial
S}\log G_t(S_1^{(j)},u)\bigg)^2\rra_{(t)}\le C_1-C_2
\lla\log G_t(S_1^{(j)},u)\rra_{(0)}.
$$
Now, since $\la\dots\ra_{(0)}$ does not depend on
$\{\xmo\}_{\mu=1}^p$, (\ref{p6.3}) follows immediately.

\end{document}